\documentclass[aps, prl, preprint]{revtex4-2}

\usepackage{bm, amsmath, amsfonts, mathtools, physics}
\usepackage[pdftex]{graphicx}
\usepackage{ulem}
\usepackage{here}

\begin{document}

  \title{Oscillatory and chaotic pattern dynamics driven by surface curvature}

  \author{Ryosuke Nishide${}^{1,}$}
    \email{r-nishide2018@g.ecc.u-tokyo.ac.jp}

  \author{Shuji Ishihara${}^{1,2}$}

  \affiliation{
  ${}^1$~Graduate School of Arts and Sciences, The University of Tokyo, Komaba 3-8-1, Meguro-ku, Tokyo 153-8902, Japan \\
  ${}^2$~Universal Biology Institute, The University of Tokyo, Komaba 3-8-1, Meguro-ku, Tokyo 153-8902, Japan
  }

\begin{abstract}
  Patterns on curved surfaces are ubiquitous,
  yet the influence of surface geometry on pattern dynamics remains elusive.
  We recently reported a new mechanism of pattern propagation in which a static pattern on a flat plane becomes a propagating pattern on a curved surface [Nishide and Ishihara, Phys. Rev. Lett. 2022].
  Here, we address whether surface curvature can drive more complex pattern dynamics beyond propagation.
  By employing a combination of weakly nonlinear analysis and numerical simulation,
  we show that oscillatory and chaotic pattern dynamics can emerge by controlling the surface shapes.
  These findings highlight a new role of surface topography in pattern formation and dynamics.
\end{abstract}

\maketitle

Surface geometry and topology have been of great interest because of their critical effects on pattern formation and dynamics.
For vector and nematic fields on closed surfaces,
the number of defects is constrained by the topology of the surface~\cite{brasselet2009vector},
and the defect dynamics were investigated in liquid crystals~\cite{kralj2011curvature, carenza2022cholesteric},
flocking~\cite{shankar2017topological},
and active nematics~\cite{keber2014topology}.
For excitable waves,
geodesic curvature modulates the wavefront and causes splitting~\cite{horibe2019curved} and swirling~\cite{gomatam1997reaction, yagisita1998spiral, McGuire2021geographic}.
One-way propagation of the excitable wave
was realized on the surface of deformed cylinders~\cite{davydov2000ring-shaped}.
In biological systems,
curved surfaces affect the sub-cellular and cellular dynamics
and can play functional roles~\cite{Baptista2019Overlooked, Ehrig2019surface, Schamberger2023curvature};
for instance,
the localization of the mechanosensitive protein Piezo1 is regulated by membrane curvature~\cite{yang2022membrane}
and cellular migration is guided by the curvature of a substrate~\cite{pieuchot2018curvotaxis}.

Turing patterns on surfaces have also received considerable attention since Turing's seminal paper~\cite{turing1952chemical},
in which he already discussed a pattern on a sphere.
Past studies revealed that the profiles and positions of the pattern change depending on the surface shapes,
such as on spheres~\cite{turing1952chemical, varea1999turing, matthews2003pattern, nunez2017diffusion,
lacitignola2017turing, sanchez2019turing},
hemispheres~\cite{liaw2001turing, nagata2013reaction},
tori~\cite{nampoothiri2017role, sanchez2019turing},
ellipsoids~\cite{nampoothiri2017role, nampoothiri2019effect},
and deformed axisymmetric cylinders and spheres~\cite{frank2019pinning}.
Notably,
these studies all presumed that a Turing pattern,
which is static on a flat plane,
remains static irrespective of the surface geometry.

Contrary to this previous assumption,
we recently reported that
the Turing pattern can exhibit propagation dynamics on curved surfaces~\cite{nishide2022pattern}.
This propagation is caused by surface curvature and is impossible in one-dimensional systems where intrinsic curvature is absent.
Through numerical and theoretical analyses,
we revealed that the onset of propagation is conditioned by symmetries of the surface and pattern.
However,
our theoretical analysis relied on the assumption that patterns propagate at a constant velocity,
thus eliminating the possibility of pattern dynamics other than propagation.

To overcome this limitation,
we adopt here a weakly nonlinear analysis~\cite{Pena2001stability, Hoyle2006pattern, Cross1993pattern},
a comprehensive approach to pattern dynamics near the Turing bifurcation point.
In this Letter,
we derive amplitude equations for reaction-diffusion equations (RDEs) on curved surfaces based on the symmetry of the systems, and show that Turing patterns can exhibit rich dynamics beyond the propagation,
including oscillatory and chaotic dynamics on curved surfaces.
In the joint
paper~\cite{joint},
we provide detailed derivation and analysis of the amplitude equations,
including the reductive perturbation method for RDEs.

\begin{figure}[h]
    \includegraphics[keepaspectratio,scale=1.0]{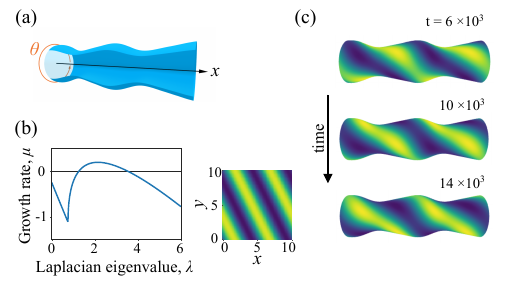}
    \caption{Propagating pattern on an axisymmetric surface.
    (a) Axisymmetric surface.
    (b) Dispersion relation and static pattern on a flat plane for Brusselator model.
    Parameters are set as $(a, b, D_u, D_v) = (2.0, 4.5, 0.5, 1.8)$.
    (c) Propagating pattern on an axisymmetric surface.
    The surface radius is $r(x)=d+k_1\cos(x)+k_2\sin(2x)$,
    where $(d,k_1,k_2) = (1.7,0.3,0.05)$ and $-2\pi \leq  x< 2\pi$.}
    \label{fig1}
  \end{figure}

\paragraph{Model and Propagating Pattern.}
  For the sake of analytical tractability~\cite{frank2019pinning,nishide2022pattern},
  we study reaction-diffusion systems on an axisymmetric surface represented as $\bm{r}=(x,r(x)\cos\theta,r(x)\sin\theta)$ with axial coordinate $x$ and the azimuth angle $\theta\in[0,2\pi)$ (Fig.~\ref{fig1}(a)) and periodic boundary conditions are used.
  We consider the general form of the RDE as follows:
  \begin{align}
    \partial_t \bm{u} = D\Delta\bm{u} + \bm{R}(\bm{u})~, \label{eq:RD}
  \end{align}
  where ${\bm u}$ is the vector representation of the chemical concentrations and is a function of the position on the surface $(x,\theta)$ and time $t$,
  $D$ is a diagonal matrix composed of the diffusion coefficients,
  $\Delta$ is the Laplace--Beltrami operator,
  and $\bm{R}(\bm{u})$ is the reaction term.
  To study the Turing pattern,
  we assume that Eq.~(\ref{eq:RD}) has a unique uniform solution ${\bm u} = {\bm u}_0$.
  The Laplace--Beltrami operator $\Delta$ characterizes the effect of the surface geometry on the system~\cite{krause2019influence}.
  For an axisymmetric surface,
  the Laplace--Beltrami operator is given by
  \begin{align}
    \label{eq:Laplace-Beltrami}
    \Delta\bullet =
            \frac{1}{r\sqrt{1+r'^2}} \partial_{x}\bigg( \frac{r}{\sqrt{1+r'^2}} \partial_{x}\bullet\bigg)
            + \frac{1}{r^2} \partial^2_{\theta}\bullet~.
  \end{align}
  For a simple cylinder,
  $\Delta$ is equivalent to the operator on a flat plane; therefore,
  the radial function $r(x)$ is not a constant function in the following analysis.
  For the analysis below,
  we will use the eigenfunction of the Laplace--Beltrami operator with respect to the eigenvalue $-\lambda$ defined by $\Delta W_{\lambda, k} = -\lambda W_{\lambda, k}$,
  which is factored into $W_{\lambda,k}(x,\theta) = X_{\lambda, k}(x)e^{ik\theta}$,
  where $k \in \mathbb{Z}$ is the wavenumber along the $\theta$-direction.
  $X_{\lambda, k}(x)$ is the solution to the following equation under the periodic boundary condition.
  \begin{align}
    \frac{r}{\sqrt{1\!+\!r'^2}}\partial_x\!\left(\!\frac{r}{\sqrt{1\!+\!r'^2}}\partial_xX_{\lambda, k}\!\right)+\left( \lambda r^2 \!-\! k^2\right)\!X_{\lambda, k}=0
  \end{align}
  The eigenvalues are two-fold degenerate for $k \neq 0$,
  for which $W_{\lambda,k}(x,\theta)$ and $W_{\lambda,-k}(x,\theta)$ are pairwise eigenfunctions that correspond to the anti-phase shift along the $\theta$-axis.

  As a representative example of the Turing pattern,
  we use the Brusselator model given by
  \begin{align}\label{eq:Brusselator}
  \begin{split}
    \partial_t u &= D_u\Delta u + u^2v - bu - u + a~, \\
    \partial_t v &= D_v\Delta v - u^2v + bu~,
    \end{split}
  \end{align}
  where, corresponding to the form of the RDE in Eq.~(\ref{eq:RD}), ${\bm u} = (u,v)^T$, $D = \textrm{diag}(D_u,D_v)$,
  and $a$ and $b$ are parameters in the reaction term $\bm{R}(\bm{u})$.
  The Brusselator model has a uniform steady solution, $\bm{u}_0 = (a, b/a)^T$.
  By choosing an appropriate parameter set, the dispersion relation,
  which is determined by the linear stability analysis at the uniform solution~\cite{nishide2022pattern},
  indicates Turing instability (Fig.~\ref{fig1}(b)),
  and the numerical simulations demonstrate the static pattern on a plane typical of Turing pattern~(Fig.~\ref{fig1}(b)).
  By contrast,
  a pattern with the same parameter exhibits propagation on an axisymmetric surface (Fig.~\ref{fig1}(c)), as shown in our previous study~\cite{nishide2022pattern}.
  Note that the dispersion relation (and thus the Turing condition)
  is the same for any surfaces, where the eigenvalue $\lambda$ matches the square of the wavenumber for a flat plane; therefore, we need to perform analysis taking into account the nonlinearity.

\paragraph{Amplitude Equations.}
  Near the Turing bifurcation point,
  the pattern has small amplitudes and can be approximated by the superposition of eigenmodes.
  Since a propagating helical pattern is approximated by two pairs of eigenfunctions with the same $k$ (see Fig.~S1 and the joint
  paper \cite{joint}),
  we investigate the dynamics of the pattern approximated by
  \begin{align}
  	 \bm{U}(t,x,\theta) &= C_{1} \bm{A}_{\lambda_1} W_{\lambda_1,k} + C_{2} \bm{A}_{\lambda_2}  W_{\lambda_2,k} + c.c.~. \label{eq:U_C1}
  \end{align}
  where c.c. stands for the complex conjugate and
  ${\bm A}_{\lambda}$ is the eigenvector of the matrix $-\lambda D+\partial_{\bm u}{\bm R}(\bm u_0)$,
  where $\partial_{\bm u}{\bm R}(\bm u_0)$ indicates the Jacobian of the reaction term at ${\bm u} = {\bm u}_0$.
  The coefficients $C_1(t)$ and $C_2(t)$ are functions of time $t$,
  and the dynamics of the pattern can be described by the amplitude equations in the form of $\partial_t C_{1} = H_{1}(C_{1},C_{2})$ and $ \partial_t C_{2} = H_{2}(C_{1}, C_{2})$.

  The symmetries inherent in a system impose strong restrictions on the possible forms of the amplitude equations.
  For an axisymmetric surface,
  the RDE is invariant about translation by arbitrary $p_\theta\in\mathbb{R}$ along the $\theta$-direction and reflection about the $\theta$-direction.
  Thus,
  the amplitude equations need to be invariant against $(C_1, C_2) \mapsto (C_1 e^{ikp_\theta}, C_2e^{ikp_\theta})$ and $ (C_1, C_2) \mapsto (\bar{C}_1, \bar{C}_2)$,
  where $\bar{C}_i$ is the complex conjugate of $C_i$.
  The general form of the amplitude equations up to the third-order expansion of $C_{i}$ is therefore given by
  \begin{align}
    \partial_tC_{1}&=a_1C_{1}+a_2|C_{1}|^2C_{1}+a_3|C_{2}|^2C_{2}+a_4|C_{1}|^2C_{2} \nonumber \\
    &\quad+a_5|C_{2}|^2C_{1}+a_6C_{1}^2\bar{C}_{2}+a_7C_{2}^2\bar{C}_{1}~, \label{eq:amp_C1}\\
    \partial_tC_{2}&=b_1C_{2}+b_2|C_{1}|^2C_{1}+b_3|C_{2}|^2C_{2}+b_4|C_{1}|^2C_{2} \nonumber \\
    &\quad+b_5|C_{2}|^2C_{1}+b_6C_{1}^2\bar{C}_{2}+b_7C_{2}^2\bar{C}_{1} ~,\label{eq:amp_C2}
  \end{align}
   where the coefficients $a_j$ and $b_j$ are real values owing to invariance against the reflection about the $\theta$-direction.
   In the following analysis,
   we assume that the coefficients are chosen such that the solution does not diverge;
   otherwise,
   higher-order terms are required to prevent divergence.
   Specific values of the coefficients can be determined from the original RDE using the reductive perturbation method (see the joint
   paper for details~\cite{joint}) yet are not required for the discussion
   in this Letter.

   By using $C_{1} = \eta e^{i\phi}$ and $C_{2}=\xi e^{i\psi}$,
   where the amplitudes $\eta$ and $\xi$ and phase variables $\phi$ and $\psi$
   are real-valued functions of time $t$,
   Eq.~(\ref{eq:U_C1}) can be written as
   $\bm{U}(t,x,\theta) = 2\eta\bm{A}_{\lambda_1}X_{\lambda_1,k}\cos(k\theta+\phi)
   +2\xi\bm{A}_{\lambda_2}X_{\lambda_2,k}\cos(k\theta+\psi)$
   and the amplitude equations are
   \begin{align}
    \partial_t\eta &= \big(a_1+a_2\eta^2+a_5\xi^2\big)\eta + a_7\eta\xi^2\cos(2\alpha) \nonumber \\
    &\quad+\big(a_3\xi^2+(a_4+a_6)\eta^2\big)\xi\cos(\alpha)~, \label{eq:eta_general} \\
    \partial_t\xi &= \big(b_1+b_3\xi^2+b_4\eta^2\big)\xi + b_6\eta^2\xi\cos(2\alpha) \nonumber \\
    &\quad+\big(b_2\eta^2+(b_5+b_7)\xi^2\big)\eta\cos(\alpha)~, \label{eq:xi_general}\\
    \eta\xi \partial_t\alpha &= -\big(a_3\xi^4+(a_4-a_6+b_5-b_7)\eta^2\xi^2+b_2\eta^4 \nonumber\\
    &\qquad\quad+2(a_7\xi^2+b_6\eta^2)\eta\xi\cos(\alpha)\big)\sin(\alpha)~, \label{eq:alpha_general}
   \end{align}
   where $\alpha \equiv \phi-\psi$ is the phase difference.
   These equations are closed in three variables---$\eta$, $\xi$, and $\alpha$---because of the translational invariance along the $\theta$-axis.
   The time evolution of $\phi$ and $\psi$ are then determined by $\eta$, $\xi$, and $\alpha$:
   \begin{align}
    \eta\partial_t\phi&=-\xi\left((a_4-a_6)\eta^2+a_3\xi^2\right)\sin(\alpha) \nonumber \\
      &\quad\quad-a_7\eta\xi^2\sin(2\alpha),\label{eq:phi_general}\\
    \xi\partial_t\psi&=+\eta\left(b_2\eta^2+(b_5-b_7)\xi^2\right)\sin(\alpha) \nonumber \\
    &\quad\quad+b_6\eta^2\xi\sin(2\alpha).\label{eq:psi_general}
   \end{align}
  The fixed-points of Eqs.~(\ref{eq:eta_general})-(\ref{eq:alpha_general}) correspond to the static and propagating patterns
  satisfying $\partial_t \phi = \partial_t \psi = 0$ and $\neq 0$, respectively.
  Thus,
  the fixed-point analysis provides an alternative manner to the previous study for determining the conditions for the
  onset of pattern propagation.
  Note that certain coefficients of the amplitude equations vanish when the surface and pattern have additional symmetry;
  for example, a helical pattern on a reflection-symmetric surface holds the inversion symmetry $\bm{U}(x,\theta) = \bm{U}(-x,-\theta)$,
  in which several interaction terms are eliminated to satisfy the symmetry as $a_3 = a_4 = a_6 = b_2 = b_5 = b_7 = 0$~\cite{joint}.
  As a result, $\partial_t \phi$ and $\partial_t \psi$ are always zero and no pattern propagation occurs.
  This line of analysis not only derives the same conditions of pattern propagation determined in the previous study~\cite{nishide2022pattern} but also identifies a non-trivial propagating pattern solution.
  See the joint
  paper for a detailed analysis~\cite{joint}.

\paragraph{Oscillatory and Chaotic Solutions in the Amplitude Equations.}
  Possible dynamics other than propagation can be investigated by searching the non-steady state solutions of Eqs.~(\ref{eq:eta_general})-(\ref{eq:alpha_general}).
  We numerically explored such dynamics using two methods,
  both of which are advantageous in that the amplitude equations allow for faster pattern exploration than solving the original Brusselator model.

\begin{figure}[h]  \includegraphics[keepaspectratio,scale=1.0]{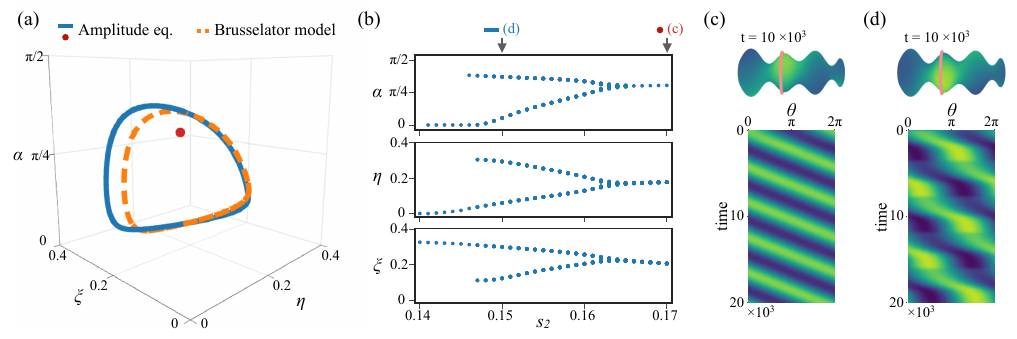}
  \caption{
  Limit cycle solution and bifurcation diagram of the amplitude equations.
  (a)~Trajectories of the limit cycle and steady-state solutions.
  Surface is chosen as $r(x) = 1 + s_2 \sin(2x) + 0.4 \sin(3x)$. Model parameters are $(a, b, D_u, D_v) = (1.505, 2.986, 0.5, 2.2)$.
  The red point and blue line represent a steady-state ($s_2 = 0.17$) and limit cycle solutions ($s_2 = 0.15$) of~Eqs.~(\ref{eq:eta_general})-(\ref{eq:alpha_general}), respectively.
  The orange dashed line represents the trajectory obtained from the original Brusselator model at $s_2 = 0.15$ (see Fig.~\ref{fig3}).
  (b)~Bifurcation diagram with respect to the surface shape parameter $s_2$. Maximal and minimal values of $\alpha, \eta, \xi$ are plotted.
  (c,d)~Propagating pattern at $s_2 = 0.17$ (c)
  and oscillatory pattern at $s_2 = 0.15$ (d).
  Kymographs along the $\theta$-axis represented by the pale red line are shown.
  See also Movie~1~\cite{supp}.
  }
  \label{fig2}
  \end{figure}

  First,
  we examined the amplitude equations derived from the Brusselator by applying the reductive perturbation method in which the coefficients are determined by the surface shape and model parameters of the original Brusselator
  near the Turing bifurcation~\cite{joint}.
  Fig.~\ref{fig2}(a) shows an example of the limit cycle solution obtained on the surface $r(x) = 1 + s_2 \sin(2x) + 0.4\sin(3x)$ with $s_2 = 0.15$
   (the coefficients of the amplitude equations are listed in Supplemental Table~S1~\cite{supp}).
   The bifurcation diagram with respect to the surface parameter $s_2$ is shown in Fig.~\ref{fig2}(b).
  At $s_2 = 0.17$, the amplitude equation reaches a fixed point corresponding to a propagating pattern (Fig.~\ref{fig2}(c)).
  As $s_2$ decreases,
  the system undergoes a Hopf bifurcation around $s_2 = 0.163$, below which
  the pattern propagates with periodic modulations and the propagation velocity is not constant
  (see the kymograph of the pattern at $s_2 = 0.15$ in Fig.~\ref{fig2}(d) and Movie~1),
  since both the amplitudes and the phase difference vary in time.
  As $s_2$ decreases further,
  the system suddenly ceases to oscillate and changes into a fixed point solution corresponding to a static pattern ($\alpha = 0$ and $\partial_t \phi = \partial_t \psi = 0$).

\begin{figure}[h]  \includegraphics[keepaspectratio,scale=1.0]
  {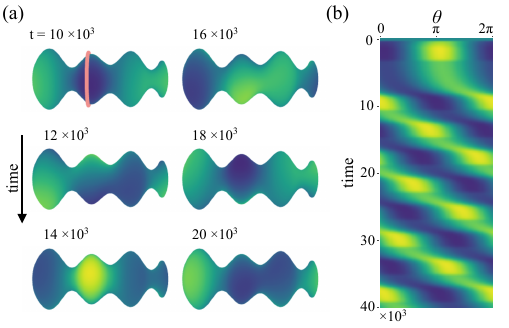}
  \caption{
  Limit cycle solution for Brusselator model.
  (a,b)~Snapshots of the limit cycle solution on an axisymmetric surface (a) and kymograph along the $\theta$-axis, represented by the pale red line (b).
  The corresponding trajectory mapped on $(\eta, \xi, \alpha)$ space is shown by the dashed orange line in Fig.~\ref{fig2}(a).
  See also Movie~1~\cite{supp}.}
  \label{fig3}
  \end{figure}

  We also confirmed that the original Brusselator model exhibits periodic dynamics similar to those obtained by the amplitude equations with parameters determined by the reductive perturbation~\cite{joint}.
  Fig.~\ref{fig3}(a) shows the trajectory of the Brusselator model in $(\eta,\xi,\alpha)$ space (orange dashed line) close to the trajectory of the amplitude equations (blue line).
  Direct observation of the pattern dynamics and kymograph (Fig.~\ref{fig3})
  further supports that the Brusselator model shows a propagating pattern with periodic modulations
  similar to that of the amplitude equations.
  Taken together,
  this example demonstrates that the Turing pattern can exhibit oscillatory dynamics other than propagation on curved surfaces.

\begin{figure}[h]
  \includegraphics[keepaspectratio,scale=1.0]{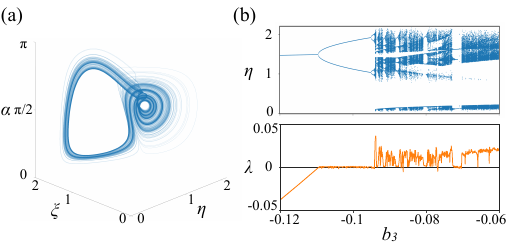}
  \caption{Chaotic dynamics in the amplitude equations.
  (a)~Trajectory of a chaotic solution ($b_3=-0.09$; see Table~S1 for the other parameters).
  (b)~Bifurcation diagram (top) and Lyapunov exponent (bottom) against $b_3$. For the bifurcation diagram,
  $\eta$ on a Poincar{\'e} section determined by $d\eta/dt = 0$ is plotted for each value of $b_3$.}
  \label{fig4}
  \end{figure}

  Next,
  we investigated the pattern dynamics by changing the coefficients of the amplitude equations $a_j$ and $b_j$ ($j=1,\ldots,7$) as free parameters
  without considering whether the parameters could be derived or not from the original RDEs.
  This search revealed that the amplitude equations can show chaotic dynamics (Fig.~\ref{fig4}(a);
  see Fig.~S2 and Table~S1~\cite{supp}
  for other examples and coefficients.)
  For the example shown in Fig.~\ref{fig4}(a),
  the appearance of the chaotic dynamics is understood in terms of period-doubling bifurcation (Fig.~\ref{fig4}(b)).
  These results suggest that the surface curvature gives rise to chaotic dynamics in a Turing pattern on curved surfaces.
  However, due to the difficulty in determining the surface and model parameters of the original RDEs from the coefficients of the amplitude equations,
  we did not find the corresponding pattern in the Brusselator model.

 \paragraph{Chaotic Pattern Dynamics.}
  The fact that chaotic solutions in the RDEs were not found in the search suggests that the parameter region for chaotic dynamics is narrow near the Turing bifurcation point and on axisymmetric surfaces.
  Therefore,
  we numerically explored the chaotic dynamics of the Brusselator model over a wider parameter range while maintaining Turing instability.

\begin{figure}[h]\includegraphics[keepaspectratio,scale=1.0]
  {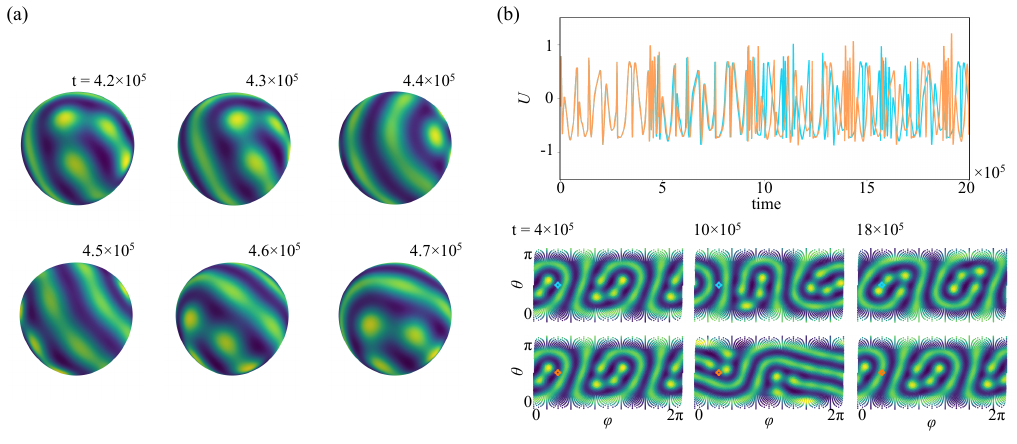}
  \caption{Chaotic dynamics of Turing pattern on a curved surface.
  (a)~Patterns on a deformed sphere.
  The surface is given by the radial function $r(\theta, \phi) = d + k (\cos (2\theta)-1) \cos(\theta) \cos(m\phi)$ in polar coordinate with $(d, k, m) = (6,0.206,2)$
  (see Supplemental Figure~S3 \cite{supp}).
  (b)~Time dependencies of two patterns with slightly different initial conditions.
  Upper panel: Time series of concentrations $U=u-a$ at the marked location in the lower panels.
  Lower panels: Projected patterns for the two trajectories at three-time points.
  The colors of the marks correspond to those of the lines in the upper panel.
  }
  \label{fig5}
  \end{figure}

  An example of a chaotic pattern found on a deformed spherical surface is shown in Fig.~\ref{fig5} and Movie~2.
  In this example,
  the pattern repeatedly takes similar but non-identical shapes while chaotically moving across the surface.
  To confirm the initial condition sensitivity of the dynamics,
  we compared two trajectories with slightly different initial conditions
  (the difference in initial conditions is of the order of $10^{-6}$ at each spatial point in the numerical simulation).
  The difference grows with time and becomes significant (Fig.~\ref{fig5}(b); see patterns of each trajectory at $t = 10 \times 10^5$ and $18 \times 10^5$ in the lower panels of the figure), clearly indicating that the dynamics are chaotic.
  In this example,
  the model parameters were chosen to be the same as those used in Fig.~\ref{fig1}.
  Therefore,
  the chaotic pattern dynamics arise due to the surface geometry.

  \paragraph{Summary.}
  In conclusion, we showed that the Turing pattern, which is static on a plane,
  can exhibit not only propagation but also more complex behaviors including oscillation and chaos on curved surfaces.
  These dynamics are switched by alternating the shape of the surface,
  indicating a new role of surface geometry for the pattern dynamics.
  Our findings hold potential significance for understanding natural phenomena,
  especially biological processes such as organ growth and cellular membrane deformation.
  For example, in animal and plant development,
  organ shape changes due to growth alter the pattern dynamics of signaling molecules on the surface,
  which feed back into the organ growth to control the eventual morphology of the organ.
  In mechano-chemical engineering,
  the interplay between chemical dynamics and surface topography guides the design and control of various processes used, for instance, in soft robotics~\cite{miller2018geometry}.
  Our findings would also have applications for manipulating chemical reactions based on the geometric characteristics of surfaces.

  In addition,
  we derived the amplitude equations for pattern dynamics on curved surfaces.
  These equations successfully capture the dynamics observed in the numerical simulations of RDEs.
  The derived equations include intricate interaction terms among modes that underly the emergence of various pattern dynamics.
  The analysis of the amplitude equations provides valuable insights into pattern dynamics on curved surfaces such as the identification of conditions for pattern propagation.
  We discussed that several interaction terms are eliminated when the surface and pattern enjoy symmetry
  such as reflection symmetry along the $x$-axis that suppresses the propagating dynamics
  (see the joint
  paper for comprehensive analysis~\cite{joint}).
  This approach using amplitude equations is useful for the control of pattern dynamics by surface shapes and will be applicable to Turing patterns on non-axisymmetric surfaces and non-Turing patterns~\cite{wittkowski2014scalar, avery2021keller},
  as well as for dynamics on network systems~\cite{nakao2010turing, vanderKolk2023emergence}
  and pattern dynamics coupled with surfaces deformation~\cite{toole2013turing, miller2018geometry, tamemoto2020pattern}.

  This study was supported by
  JST SPRING JPMJSP2108 (to R.N.),
  JSPS KAKENHI JP23KJ0641 (to R.N.),
  JPJSJPR 20191501 (to S.I.),
  and JST CREST JPMJCR1923, Japan (to S.I.).

  R.N. and S.I. both proposed the research direction,
  contributed to the theoretical analysis,
  and wrote the manuscript.
  R.N. performed all numerical simulations.

\end{document}